\documentclass[aps, pre, reprint, longbibliography]{revtex4-1} 
\usepackage[dvipdfmx]{graphicx}
\usepackage{dcolumn}
\usepackage{amsmath}
\usepackage{amssymb}
\usepackage{color}
\usepackage{bm}
\usepackage{txfonts}

\begin{document}
\title{Interscale entanglement production in a quantum system simulating classical chaos}
\author{Taiki Haga}
\email[]{taiki.haga@omu.ac.jp}
\affiliation{Department of Physics and Electronics, Osaka Metropolitan University, Osaka, 599-8531, Japan}
\author{Shin-ichi Sasa}
\affiliation{Department of Physics, Kyoto University, Kyoto, 606-8502, Japan}
\date{\today}

\begin{abstract}
It is a fundamental problem how the universal concept of classical chaos emerges from the microscopic description of quantum mechanics.
We here study standard classical chaos in a framework of quantum mechanics.
In particular, we design a quantum lattice system that exactly simulates classical chaos after an appropriate continuum limit, which is called the ``Hamiltonian equation limit''.
The key concept of our analysis is an entanglement entropy defined by dividing the lattice into many blocks of equal size and tracing out the degrees of freedom within each block.
We refer to this entropy as the ``interscale entanglement entropy'' because it measures the amount of entanglement between the microscopic degrees of freedom within each block and the macroscopic degrees of freedom that define the large-scale structure of the wave function.
By numerically simulating a quantum lattice system corresponding to the Hamiltonian of the kicked rotor, we find that the long-time average of the interscale entanglement entropy becomes positive only when chaos emerges in the Hamiltonian equation limit, and the growth rate of the entropy in the initial stage is proportional to that of the coarse-grained Gibbs entropy of the corresponding classical system.
\end{abstract}

\maketitle

\section{Introduction}

All complex dynamics in our daily life should be described by the quantum mechanics of atoms and light.
Nevertheless, clear physical concepts such as thermodynamics, information, computation, and chaos are formulated for macroscopic natural phenomena without respecting the quantum mechanics of such microscopic degrees of freedom.
The fundamental problem here is how these macroscopically universal concepts emerge from quantum mechanics.
While the microscopic basis of thermodynamics has been established as equilibrium statistical mechanics, the characterization of information and chaos from the perspective of microscopic physics still remains explored \cite{Parrondo-15, Goold-16, Evans-02, Seifert-12, Zurek-03, Gemmer, Nielsen, Guhr-98, Alhassid-00, Gutzwiller, Haake}.

A major problem in the characterization of chaos by quantum mechanics is that the nature of irregularities in quantum mechanics is quite different from that in classical mechanics.
Chaos in classical mechanics is due to the exponential sensitivity of trajectories with respect to initial conditions.
The complexity of dynamics is characterized by the Kolmogorov-Sinai (KS) entropy $h_{\mathrm{KS}}$, which is defined as the rate of increase of the Shannon entropy corresponding to the probability of an ensemble of trajectories \cite{Eckmann-85, Boffetta-02} (see Appendix~\ref{appendix:KS_entropy} for the definition of $h_{\mathrm{KS}}$). 
On the other hand, the time evolution of any quantum state is represented by a superposition of periodically oscillating energy eigenstates.
This implies that the dynamics of a quantum system with a finite number of levels is necessarily quasi-periodic.
In quantum systems that exhibit chaos in the classical limit, each energy eigenstate is known to have an irregular spatial structure \cite{Haake}, and hence the time evolution of a quantum state represented by a superposition of many such eigenstates can be highly irregular in a different sense than in classical systems.
Since classical mechanics is believed to emerge from quantum mechanics under an appropriate limit, the irregularities of classical chaos should be closely related to the irregularities inherent in quantum mechanics.
In particular, the KS entropy of a classical chaotic system should be determined from the time evolution of the wave function of the corresponding quantum system.

The dynamical generation of entanglement entropy in quantum systems that exhibit chaos in the classical limit has received much attention.
In quantum chaotic systems with two or more degrees of freedom, if the initial state is taken to be a product state, the entanglement entropy $S$ between subsystems begins to increase linearly with time, $S(t) \sim ht$, and eventually saturates at some equilibrium value.
The relation between the initial growth rate $h$ of the entanglement entropy and the KS entropy $h_{\mathrm{KS}}$ of the corresponding classical system has been investigated over the past few decades \cite{Zurek-94, Zarum-98, Miller-99-1, Miller-99-2, Pattanayak-99, Monteoliva-00, Monteoliva-01, Demkowicz-04, Asplund-16, Bianchi-18, Tanaka-02, Fujisaki-03, Jacquod-04, Jacquod-09}.
Since the KS entropy $h_{\mathrm{KS}}$ is presumed to be equal to the growth rate of the classical Gibbs entropy corresponding to the coarse-grained probability distribution in phase space (see Appendix~\ref{appendix:Gibbs_entropy} or Refs.~\cite{Latora-99} and \cite{Vulpiani-05}), it is natural to expect a strong correlation between the entanglement growth rate $h$ and $h_{\mathrm{KS}}$ of the classical system.
In fact, early works by Miller and Sarkar \cite{Miller-99-1, Miller-99-2} suggested that the rate of entropy generation increases linearly with $h_{\mathrm{KS}}$.
However, Tanaka {\it et al.} \cite{Tanaka-02, Fujisaki-03} pointed out that if the coupling between subsystems is sufficiently weak, increasing the strength of chaos does not enhance the rate of entanglement generation.
Jacquod and Petitjean \cite{Jacquod-04, Jacquod-09} also argued that whether the rate of entropy generation is given by the KS entropy depends on the details of the interaction between subsystems. 
Therefore, the rate of entanglement generation between subsystems is not always a universal measure of chaos.

The central object of this study is quantum entanglement between microscopic and macroscopic degrees of freedom.
In classical chaotic systems, microscopic details in phase space expand into macroscopic structures as the system evolves over time.
In other words, there is a flow of information from microscopic to macroscopic scales, the amount of which is the KS entropy $h_{\mathrm{KS}}$.
Therefore, when trying to understand the quantum mechanical origin of chaos, it is natural to consider entanglement between degrees of freedom at different scales.
This is clearly different from the entanglement between subsystems of a bipartite system that has been studied in previous works.
Such ``interscale entanglement" is expected to be positive if the system exhibits chaos in the classical limit and zero if it exhibits regular behavior.

To define a measure of the interscale entanglement, we introduce a tight-binding model that describes the quantum dynamics of a single particle on a lattice.
As a remarkable property of this model, we can take two kinds of continuum limit.
One is the ``Schr\"odinger equation limit'', in which the time evolution of the continuous wave function obeys the standard Schr\"odinger equation.
The other is the ``Hamiltonian equation limit'', in which a localized wave function obeys the Hamiltonian equation.
Therefore, we call this model a unified simulator of the Schr\"odinger equation and classical Hamiltonian equation  (see Fig.~\ref{fig-tight-binding}).
For this quantum lattice system, we define the ``interscale entanglement entropy (IEE)" by a simple block-spin coarse-graining procedure.
Namely, we divide the tight-binding model into many blocks of equal size and calculate the von-Neumann entropy of the reduced density matrix obtained by tracing out degrees of freedom within each block.
The entanglement entropy defined here measures the amount of entanglement between microscopic degrees of freedom within each block and macroscopic degrees of freedom that define the large-scale structure of the wave function.

We then attempt to characterize chaos emerging in the lattice model in terms of the IEE.
Two natural questions arise here.
First, how does the behavior of the IEE in early time relate to the KS entropy in the Hamiltonian equation limit? 
It is natural to expect a strong correlation between the initial growth rate of the IEE and the KS entropy.
Second, what happens when the Hamiltonian equation limit is taken after the long-time limit? 
Note that different behavior is expected depending on the order of the two limits.
Since there are no clear trajectories in the long time limit before taking the Hamiltonian equation limit, the system obtained in this manner may be related to an ensemble description of classical chaos.
Thus, the long-time behavior of the IEE can provide another characterization of chaos.

We employ the kicked rotor (or the standard map) as a prototypical model of classical chaos.
We first study the case that the long-time limit is taken first, and then  the Hamiltonian equation limit is considered.
We find that, while the long-time average of the IEE vanishes in the Hamiltonian equation limit when dynamics is regular, it is positive when chaos emerges.
We next study the case that the Hamiltonian equation limit is taken first, and then the long-time limit is considered.
In particular, we focus on the initial growth rate of the IEE.
Starting from a well-localized wave packet, we observe a linear growth of IEE in early time.
We expect that in the Hamiltonian equation limit this linear growth continues persistently without saturation.
We then observe that the growth rate of the IEE is proportional to that of the classical Gibbs entropy.
These observations confirm that the IEE defined here has desirable properties as a measure of entanglement between different scales.

This paper is organized as follows.
In Sec.~\ref{sec:unified_simulator}, we define a tight-binding model on a lattice.
This model can simulate either the Schr\"odinger equation or the classical Hamiltonian equation under an appropriate continuum limit.
In Sec.~\ref{sec:interscale_entanglement_entropy}, we introduce the IEE by a coarse-graining procedure in a discrete configuration space.
In Sec.~\ref{sec:kicked_rotor}, we numerically demonstrate that for the kicked rotor, the long-time average of the IEE has a nonzero value for chaotic initial conditions, and that the growth rate of the IEE is proportional to that of the classical Gibbs entropy.
Finally, Sec.~\ref{sec:conclusions} is devoted to conclusions and discussions.
In Appendix~\ref{appendix:KS_entropy}, we briefly review the definition of the KS entropy for classical systems.
The connection to the growth rate of the Gibbs entropy is also discussed in Appendix~\ref{appendix:Gibbs_entropy}.
Appendix \ref{appendix:Hamiltonian_equation_limit} provides a detailed discussion of the Hamiltonian equation limit in the tight-binding model.
In Appendix \ref{appendix:different_beta}, we present numerical data for different values of the scaling exponent in the Hamiltonian equation limit.

\section{Unified simulator of quantum and classical dynamics}
\label{sec:unified_simulator}

\begin{figure}
	\centering
	\includegraphics[width=0.45\textwidth]{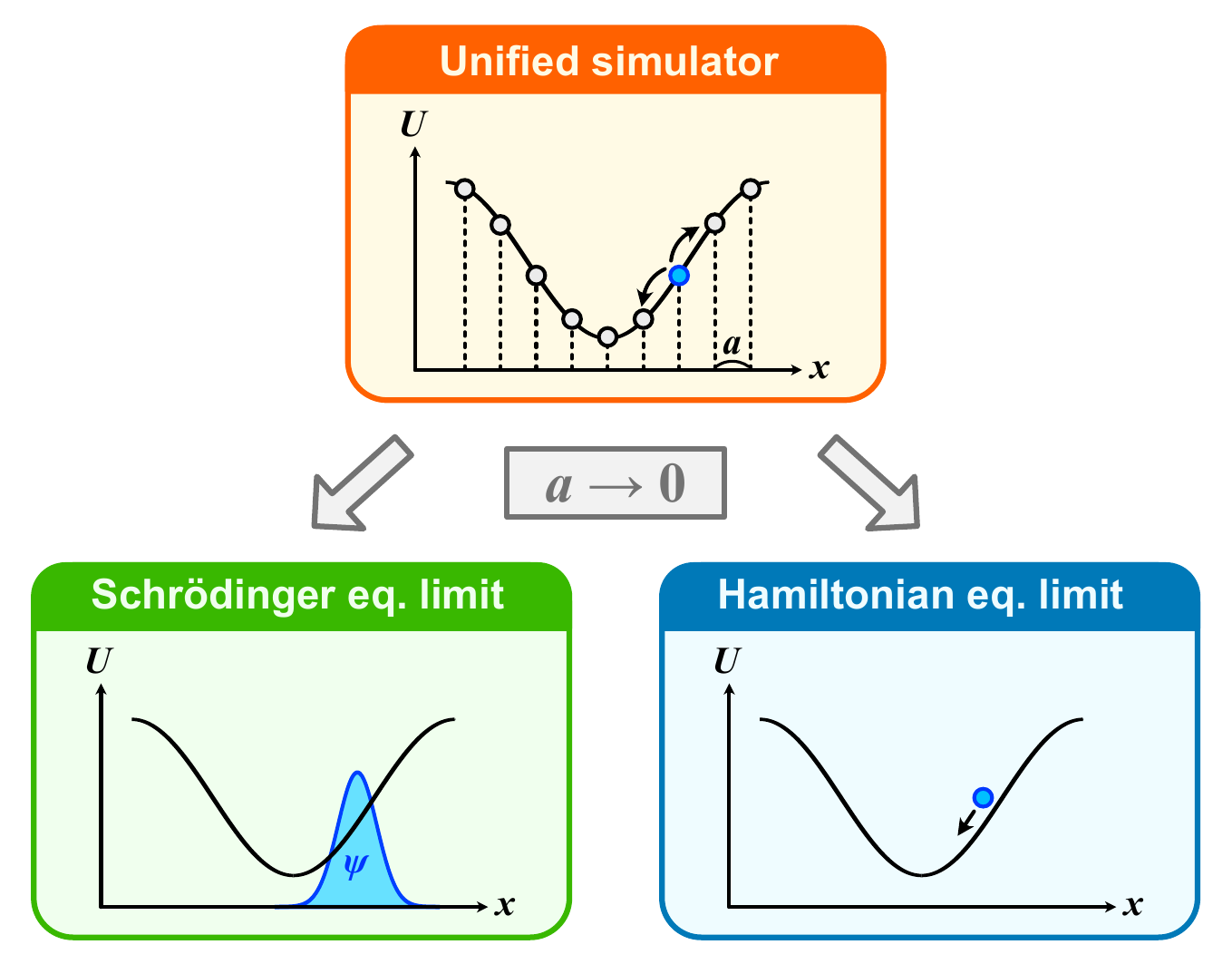}
	\caption{Schematic illustration of the unified simulator.
		The tight-binding model describing the dynamics of a particle under potential $U(x)$ has two continuum limits, the Schr\"odinger equation limit and the Hamiltonian equation limit.}
	\label{fig-tight-binding}
\end{figure}

Let us consider a tight-binding model describing the dynamics of a particle on a one-dimensional lattice.
The Hamiltonian is given by
\begin{equation}
	\hat{H} = -J \sum_n (\hat{a}_{n+1}^{\dag} \hat{a}_n + \hat{a}_{n}^{\dag} \hat{a}_{n+1}) + g \sum_n U(x_n;t) \hat{a}_{n}^{\dag} \hat{a}_{n},
	\label{General_Hamiltonian}
\end{equation}
where $\hat{a}_n^{\dag}$ and $\hat{a}_n$ are the creation and annihilation operators of a boson at site $n$, which satisfy the commutation relation $[\hat{a}_m,\hat{a}_n]=[\hat{a}_m^{\dag},\hat{a}_n^{\dag}]=0$ and $[\hat{a}_m,\hat{a}_n^{\dag}]=\delta_{mn}$, and $U(x;t)$ is a time-dependent potential.
The parameters $J$ and $g$ represent the tunneling amplitude and potential strength, respectively.
We denote the position of site $n$ as $x_n=na$ in terms of the lattice constant $a$.
If we denote the state in which the particle is located at site $n$ as $|n \rangle$, a state vector is represented as
\begin{equation}
	|\psi \rangle = \sum_n \psi_n |n \rangle,
\end{equation}
where $\psi_n$ is the wave function.
The time evolution of the state vector is described by
\begin{equation}
	i \hbar \partial_t |\psi \rangle = \hat{H} |\psi \rangle.
	\label{Schrodinger}
\end{equation}
In appropriate continuum limits, this lattice model provides a unified simulator for the Schr\"{o}dinger equation and the Hamiltonian equation (see Fig.~\ref{fig-tight-binding}).
In the following, we omit the time variable ``$t$'' in the potential $U(x;t)$ for simplicity.

It is not difficult to see that Eq.~(\ref{Schrodinger}) reduces to the standard Schr\"{o}dinger equation in an appropriate continuum limit $a \to 0$.
Since $\langle n| \hat{H} |n \rangle = g U(x_n)$ and $\langle n| \hat{H} |n\pm1 \rangle = -J$, Eq.~(\ref{Schrodinger}) is rewritten as
\begin{equation}
	i \hbar \partial_t \psi_n = -J (\psi_{n+1}-2\psi_n+\psi_{n-1}) + g U(x_n) \psi_n,
\end{equation}
where we have added to $\hat{H}$ a constant $2J$ to ensure a well-defined continuum limit.
In the limit $a \to 0$ with $Ja^2=\hbar^2/2m$ and $g=1$, the time evolution of $\psi(x_n)=\psi_n$ is described by
\begin{equation}
	i \hbar \partial_t \psi(x) = \left[ -\frac{\hbar^2}{2m} \partial_x^2 + U(x) \right] \psi(x),
	\label{Schrodinger_wave_func}
\end{equation}
where $m$ is the mass of the particle.
We call the above limit the Schr\"odinger equation limit.

We can also show that in another continuum limit, the time evolution of a localized wave packet is described by the classical equation of motion with the Hamiltonian
\begin{equation}
	H_{\mathrm{cl}} = \frac{p^2}{2m} + U(x).
	\label{classical_Hamiltonian}
\end{equation}
We refer to such a limit as the Hamiltonian equation limit.
Since the discussion involves a reformulation of the well-known Ehrenfest theorem for the tight-binding model, we summarize the results in this section and present the details in Appendix \ref{appendix:Hamiltonian_equation_limit}.

The operators of position $\hat{x}$ and momentum $\hat{p}$ of the particle are defined by
\begin{equation}
	\hat{x} := \sum_n x_n \hat{a}_{n}^{\dag} \hat{a}_{n},
	\label{def_x_hat}
\end{equation}
in terms of $x_n = na$, and
\begin{equation}
	\hat{p} := i\frac{mJa}{\hbar} \sum_n (\hat{a}_{n+1}^{\dag} \hat{a}_n - \hat{a}_{n}^{\dag} \hat{a}_{n+1}),
	\label{def_p_hat}
\end{equation}
where $m$ is the mass in Eq.~\eqref{classical_Hamiltonian}.
The characteristic length scale of the potential is given by
\begin{equation}
	l_{U} := \frac{U_{\mathrm{max}}-U_{\mathrm{min}}}{\max_{x} |U'(x)|},
\end{equation}
where $U_{\mathrm{max}}$ and $U_{\mathrm{min}}$ are the maximal and minimal values of $U(x)$, respectively.
We assume that the width of the wave packet $\sigma_x := (\langle \hat{x}^2 \rangle - \langle \hat{x} \rangle^2)^{1/2}$ satisfies
\begin{equation}
	a \ll \sigma_x \ll l_{U}.
	\label{classical_condition_1}
\end{equation}
Furthermore, it is also assumed that the difference in phase of the wave functions $\psi_n=|\psi_n|e^{i\theta_n}$ at adjacent sites is small,
\begin{equation}
	|\theta_{n+1}-\theta_n| \ll 1.
	\label{classical_condition_2}
\end{equation}
In other words, the conditions \eqref{classical_condition_1} and \eqref{classical_condition_2} mean that the wave function varies slowly compared to the lattice constant, but is localized compared to $l_U$.
Then, the expectation values of $\hat{x}$ and $\hat{p}$ satisfy
\begin{equation}
	m \frac{d}{dt} \langle \hat{x} \rangle = \langle \hat{p} \rangle,
	\label{x_EOM}
\end{equation}
\begin{equation}
	\frac{d}{dt} \langle \hat{p} \rangle \simeq -\frac{2mJga^2}{\hbar^2} U'(\langle \hat{x} \rangle),
	\label{p_EOM}
\end{equation}
which implies that $\langle \hat{x} \rangle$ and $\langle \hat{p} \rangle$ follow the classical equation of motion associated with the Hamiltonian (\ref{classical_Hamiltonian}) if
\begin{equation}
	2mJga^2 = \hbar^2.
	\label{relation_J_g}
\end{equation}

We assume that in the continuum limit $a \to 0$, the tunneling amplitude $J$ and the potential strength $g$ scale as
\begin{equation}
	J \propto a^{-1-\beta}, \quad g \propto a^{-1+\beta},
	\label{J_g_scaling}
\end{equation}
where $\beta$ is an appropriate exponent to be specified later.
In this continuum limit, the left-hand side of Eq.~(\ref{relation_J_g}) is independent of $a$.
If one chooses $\beta=1$, this limit corresponds to the Schr\"odinger equation limit.
Let us consider an initial wave packet satisfying Eqs.~(\ref{classical_condition_1}) and (\ref{classical_condition_2}).
In early time, the dynamics of $\langle \hat{x} \rangle$ and $\langle \hat{p} \rangle$ are described by the classical equation of motion (\ref{p_EOM}).
However, in later time, the trajectories of $\langle \hat{x} \rangle$ and $\langle \hat{p} \rangle$ start to deviate from the solution of the classical equation of motion as the width of the wave packet becomes comparable to the characteristic length scale $l_U$ of the potential.
Let $\tau_c$ be the timescale where either Eq.~(\ref{classical_condition_1}) or (\ref{classical_condition_2}) breaks down.
We need to determine the range of $\beta$ such that $\tau_c$ diverges to infinity as the lattice constant $a$ goes to zero.
From the argument in Appendix \ref{appendix:Hamiltonian_equation_limit}, we obtain the condition of $\beta$ for the Hamiltonian equation limit:
\begin{equation}
	0<\beta<1.
	\label{beta_CCL}
\end{equation}
If we choose an initial wave packet with the minimal uncertainty, $\tau_c$ can be estimated as
\begin{equation}
	\tau_c \sim \frac{1}{\lambda} \ln \frac{l_U}{a^{(1-\beta)/2}},
	\label{tau_c}
\end{equation}
where $\lambda$ is the largest Lyapunov exponent of the classical system.

\section{Interscale entanglement entropy}
\label{sec:interscale_entanglement_entropy}

We first recall the definition of the conventional entanglement entropy for a bipartite system.
Let $\mathcal{H}$ be the Hilbert space of the total system with dimension $D$.
We assume that the total system is composed of two subsystems, whose Hilbert spaces are $\mathcal{H}_A$ and $\mathcal{H}_B$ with dimensions $D_A$ and $D_B$, respectively.
Then, we have a decomposition of $\mathcal{H}$,
\begin{equation}
	\mathcal{H} = \mathcal{H}_A \otimes \mathcal{H}_B,
\end{equation}
where $D=D_A D_B$.
If we define orthonormal bases of $\mathcal{H}_A$ and $\mathcal{H}_B$ as $\{ |\varphi^A_{\nu} \rangle \}_{\nu=1,...,D_A}$ and $\{ |\varphi^B_n \rangle \}_{n=1,...,D_B}$, respectively, any pure state $|\psi \rangle$ of the total system is written as
\begin{equation}
	|\psi \rangle = \sum_{\nu=1}^{D_A} \sum_{n=1}^{D_B} \psi_{\nu n} |\varphi^A_{\nu} \rangle \otimes |\varphi^B_n \rangle.
\end{equation}
The reduced density matrix for $\mathcal{H}_A$ is given by
\begin{equation}
	\hat{\rho}_A = \mathrm{Tr}_B[|\psi \rangle \langle \psi|] = \sum_{\mu, \nu=1}^{D_A} \sum_{n=1}^{D_B} \psi_{\mu n} \psi_{\nu n}^* |\varphi^A_{\mu} \rangle \langle \varphi^A_{\nu}|,
\end{equation}
and then, the entanglement entropy is defined as
\begin{equation}
	S=-\mathrm{Tr}_A [\hat{\rho}_A \ln \hat{\rho}_A].
\end{equation}

\begin{figure}
	\centering
	\includegraphics[width=0.45\textwidth]{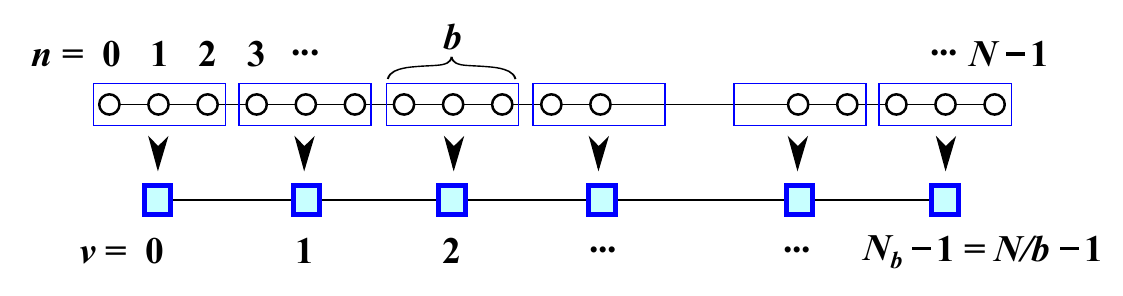}
	\caption{Schematic illustration of the coarse-graining procedure.
		The lattice with $N$ sites is divided into $N_b$ blocks with size $b$ ($b=3$ in this figure).
		The degrees of freedom within each block are traced out to obtain a coarse-grained density matrix.}
	\label{fig-coarse-graining}
\end{figure}

Next, let us introduce a different type of entanglement entropy for the tight-binding model defined in the previous section.
Here, we shall employ a simple block-spin decimation procedure.
We first divide the lattice with $N$ sites into $N_b$ blocks with size $b=N/N_b$ (see Fig.~\ref{fig-coarse-graining}).
Below, the Roman alphabets $m,n=0,...,N-1$ denote the site index of the original lattice and the Greek alphabets $\mu,\nu=0,...,N_b-1$ denote the block index.
Any $n$ can be uniquely expressed as $n=\nu b + j$ by using $\nu$ and $j$ $(0 \leq \nu < N_b, \: 0 \leq j < b)$.
We formally rewrite the state in which the particle resides at site $n$ as
\begin{equation}
	| n \rangle = | \Phi_{\nu} \rangle \otimes | \phi^{j} \rangle,
	\label{bipartite_base_block}
\end{equation}
where $\{ | \Phi_{\nu} \rangle \}_{\nu=0,...,N_b-1}$ and $\{ | \phi^{j} \rangle \}_{j=0,...,b-1}$ are orthonormal bases in Hilbert spaces $\mathcal{H}_{\Phi}$ and $\mathcal{H}_{\phi}$ with dimensions $N_b$ and $b$, respectively.
In other words, $| \Phi_{\nu} \rangle$ denotes a ``macroscopic'' state in which the particle belongs to the $\nu$ th block, and $| \phi_{j} \rangle$ denotes a ``microscopic'' state in which the particle is located at the $j$ th site in some block.
Equation (\ref{bipartite_base_block}) defines a formal decomposition of the total Hilbert space,
\begin{equation}
	\mathcal{H}=\mathcal{H}_{\Phi} \otimes \mathcal{H}_{\phi}.
\end{equation}
Any state vector can be written as
\begin{equation}
	| \psi \rangle = \sum_{n=0}^{N-1} \psi_n | n \rangle = \sum_{\nu=0}^{N_b-1} \sum_{j=0}^{b-1} \psi_{\nu}^{j} | \Phi_{\nu} \rangle \otimes | \phi^{j} \rangle.
\end{equation}
We define a coarse-grained density matrix by tracing out the degrees of freedom within each block,
\begin{equation}
	\hat{\rho}_{\Phi} = \mathrm{Tr}_{\phi} [| \psi \rangle \langle \psi |] = \sum_{\mu, \nu =0}^{N_b-1} \rho_{\mu \nu} | \Phi_{\mu} \rangle \langle \Phi_{\nu} |,
\end{equation}
where the matrix element $\rho_{\mu \nu}$ is given by
\begin{equation}
	\rho_{\mu \nu} = \sum_{j=0}^{b-1} \psi_{\mu}^j (\psi_{\nu}^j)^*.
	\label{rho_mu_nu_psi}
\end{equation}
We denote the eigenvalues of $\rho_{\mu \nu}$ as $\{ w_{\alpha} \}_{\alpha=0,...,N_b-1}$, $(0 \leq w_{\alpha} \leq 1)$, and then, the entropy is defined by
\begin{equation}
	S = - \sum_{\alpha=0}^{N_b-1} w_{\alpha} \ln w_{\alpha}.
\end{equation}

This entropy has the following properties.
When each block involves only one site $(b=1)$ or all sites are involved in a single block $(b=N)$, $S=0$ from its definition.
When the wave function $\psi_n$ varies slowly in space and it can be considered as a constant within each block $(\psi_{\nu}^{j} \simeq \psi_{\nu})$, the coarse-grained density matrix is written as
\begin{equation}
	\hat{\rho}_{\Phi} \simeq \biggl(\: \sum_{\mu=0}^{N_b-1} \sqrt{b} \psi_{\mu} | \Phi_{\mu} \rangle \biggr) \biggl(\: \sum_{\nu=0}^{N_b-1} \sqrt{b} \psi_{\nu}^* \langle \Phi_{\nu} | \biggr),
\end{equation}
which approximately describes a pure state, and thus, we conclude $S \simeq 0$.
Conversely, when the wave function is localized in a single block, for example $\mu=0$, the matrix element of $\hat{\rho}_{\Phi}$ reads $\rho_{\mu \nu} \simeq \delta_{\mu 0} \delta_{\nu 0}$, and consequently, we have $S \simeq 0$.
The maximum entropy is achieved for the ``infinite-temperature state'' $\rho_{\mu \nu}=\delta_{\mu \nu}/N_b$, which is realized when the wave function $\psi_{\mu}^j$ is completely random.
In such a case, by the central limit theorem, the off-diagonal elements of Eq.~\eqref{rho_mu_nu_psi} are suppressed by $b^{-1/2}$ compared to its diagonal elements.
One can understand these properties from the fact that $S$ is a measure of the information loss associated with the elimination of the microscopic degrees of freedom.
The vanishing of $S$ for a slowly varying state implies that there is no information loss in the coarse-graining process.

The entropy defined above measures the amount of the entanglement between the macroscopic and microscopic degrees of freedom.
Thus, we call it the interscale entanglement entropy (IEE).
Since it can be considered as a quantum analog of the coarse-grained Gibbs entropy defined in Appendix \ref{appendix:Gibbs_entropy}, we expect that the growth rate of the IEE is related to the KS entropy in the Hamiltonian equation limit.
In the next section, we numerically investigate the behavior of the IEE for the kicked rotor.

We remark on the generalization the IEE to multi-particle cases.
For an $\mathcal{N}$-particle classical system in $\mathcal{D}$ spatial dimensions, one has a $\mathcal{D}\mathcal{N}$-dimensional configuration space $(\mathbf{r}_1,...,\mathbf{r}_{\mathcal{N}})$, where $\mathbf{r}_i=(r_i^1,...,r_i^{\mathcal{D}})$ is the coordinate of the $i$ th particle.
We can define a {\it single-particle} tight-binding model on a $\mathcal{D}\mathcal{N}$-dimensional hypercubic lattice.
In an appropriate continuum limit, the time evolution of a wave packet in the tight-binding model is described by the classical equation of motion.
The IEE is defined by the similar block-spin coarse-graining procedure in the $\mathcal{D}\mathcal{N}$-dimensional configuration space.
This entropy measures the information loss associated with the coarse-graining of a many-body wave function $\psi(\mathbf{r}_1,...,\mathbf{r}_{\mathcal{N}})$.

\section{Entropy production in the kicked rotor}
\label{sec:kicked_rotor}

The Hamiltonian of the classical kicked rotor is defined by
\begin{equation}
	H(t) = \frac{1}{2} p^2 + K \cos x \sum_{\tau =-\infty}^{\infty} \delta(t-\tau),
	\label{Hamiltonian_kicked_rotor}
\end{equation}
where $K$ is the kick strength and the mass of the particle is set to unity.
We restrict the position $x$ in $[0,2\pi)$ by imposing the periodic boundary condition.
The equation of motion reads
\begin{eqnarray}
	\left\{
	\begin{array}{l}
		p_{\tau+1} = p_{\tau} + K \sin x_{\tau}, \\
		x_{\tau+1} = x_{\tau} + p_{\tau+1},
	\end{array}
	\right.
	\label{EM_classical_kicked_rotor}
\end{eqnarray}
where $x_{\tau}$ and $p_{\tau}$ denote the position and momentum of the particle at time $t=\tau \:(\tau \in \mathbb{Z})$.
The behaviors of the classical kicked rotor (also known as the standard map) are summarized as follows \cite{Lichtenberg}.
For $K=0$, we have only trivial solutions $p_t=p_0$ and $x_t=x_0+p_0 t$.
For $0<K<K_c \simeq 0.97$, since the phase space is separated by invariant tori in which the dynamics is regular, the kinetic energy $p^2/2$ remains finite.
For $K_c < K$, these global invariant tori are destroyed, and then $p^2/2$ grows linearly in time.
Finally, for $K>4$, most part of the phase space is filled by chaotic trajectories.
The KS entropy asymptotically behaves as $h_{\mathrm{KS}} \simeq \ln (K/2)$ for $K>4$.

The quantum kicked rotor model is defined by replacing $x$ and $p$ in Eq.~(\ref{Hamiltonian_kicked_rotor}) with operators $\hat{x}$ and $\hat{p}$, respectively.
In this section, we set the Planck's constant $\hbar$ to unity.
The time-evolution operator for a single period, which is also called Floquet operator, is given by
\begin{equation}
	\hat{U} = e^{-i\hat{p}^2/2} e^{-iK\cos \hat{x}}.
\end{equation}
In contrast to the classical case, the expectation value of the kinetic energy $\langle \hat{p}^2/2 \rangle$ remains finite for an arbitrary $K$, because the eigenstates of the Floquet operator $\hat{U}$ exhibit Anderson localization in the momentum space \cite{Casati-90, Izrailev-90, Tian-10}.

Following the method explained in Sec.~\ref{sec:unified_simulator}, we define a tight-binding version of the kicked rotor on a one-dimensional lattice.
The position of each site of the lattice is written as $\{ x_n = na \}_{n=0,...,N-1}$ in terms of the lattice constant $a = 2\pi/N$, where $N$ is the number of the sites.
The Hamiltonian of the tight-binding model is given by Eq.~(\ref{General_Hamiltonian}) with a time-dependent potential
\begin{equation}
	U(x_n;t)=K \cos x_n \sum_{\tau=-\infty}^{\infty} \delta(t-\tau).
\end{equation}
From Eq.~(\ref{beta_CCL}), we set
\begin{equation}
	J = a^{-3/2}, \:\:\:\: g= \frac{1}{2} a^{-1/2},
	\label{J_g_KR}
\end{equation}
which satisfy Eq.~(\ref{relation_J_g}).
Starting from an initial wave packet with the minimal uncertainty, the expectation values of the position and momentum follow Eq.~(\ref{EM_classical_kicked_rotor}) up to a timescale $\tau_{\mathrm{c}}$ given by Eq.~(\ref{tau_c}).

As an initial state, we employ the following wave function:
\begin{equation}
	\psi_n \propto \psi_n^{(\mathrm{PW})} \psi_n^{(\mathrm{EN})},
	\label{initial_psi}
\end{equation}
where the plane-wave part is given by
\begin{equation}
	\psi_n^{(\mathrm{PW})} = \exp \biggl( i \frac{\bar{p}_0 x_n}{2Ja^2} \biggr),
	\label{initial_psi_pw}
\end{equation}
and the envelope of the wave function is defined as
\begin{equation}
	\psi_n^{(\mathrm{EN})} = \exp \biggl[ - \frac{(\cos x_n - \cos \bar{x}_0)^2+(\sin x_n - \sin \bar{x}_0)^2}{2\sigma_{x0}^2} \biggr],
	\label{initial_psi_en}
\end{equation}
where $\bar{x}_0 \in [0,2\pi)$ and $\sigma_{x0}$ denote the position and width of the wave packet, respectively.
From the periodic boundary condition, we have $\bar{p}_0=2Ja^2m \ (m \in \mathbb{Z})$.
Equation (\ref{appendix:uncertainty}) in Appendix \ref{appendix:Hamiltonian_equation_limit} implies the uncertainty relation $\sigma_x \sigma_p \sim a^{1/2}$.
In order to reproduce the classical dynamics in the continuum limit, $\sigma_x$ and $\sigma_p$ must simultaneously vanish.
Here, we choose $\sigma_{x0}$ in Eq.~(\ref{initial_psi}) as
\begin{equation}
	\sigma_{x0} = a^{1/4}.
	\label{sigma_x0}
\end{equation}
One can easily confirm that the expectation value of the momentum $\langle \hat{p} \rangle = iJa \sum_n(\psi_{n+1}^* \psi_{n} - \psi_{n}^* \psi_{n+1})$ for the initial state (\ref{initial_psi}) converges to $\bar{p}_0$ in the continuum limit.

\subsection{Long-time averaged entropy}
\label{sec:long_time_average}

\begin{figure}
	\centering
	\includegraphics[width=0.45\textwidth]{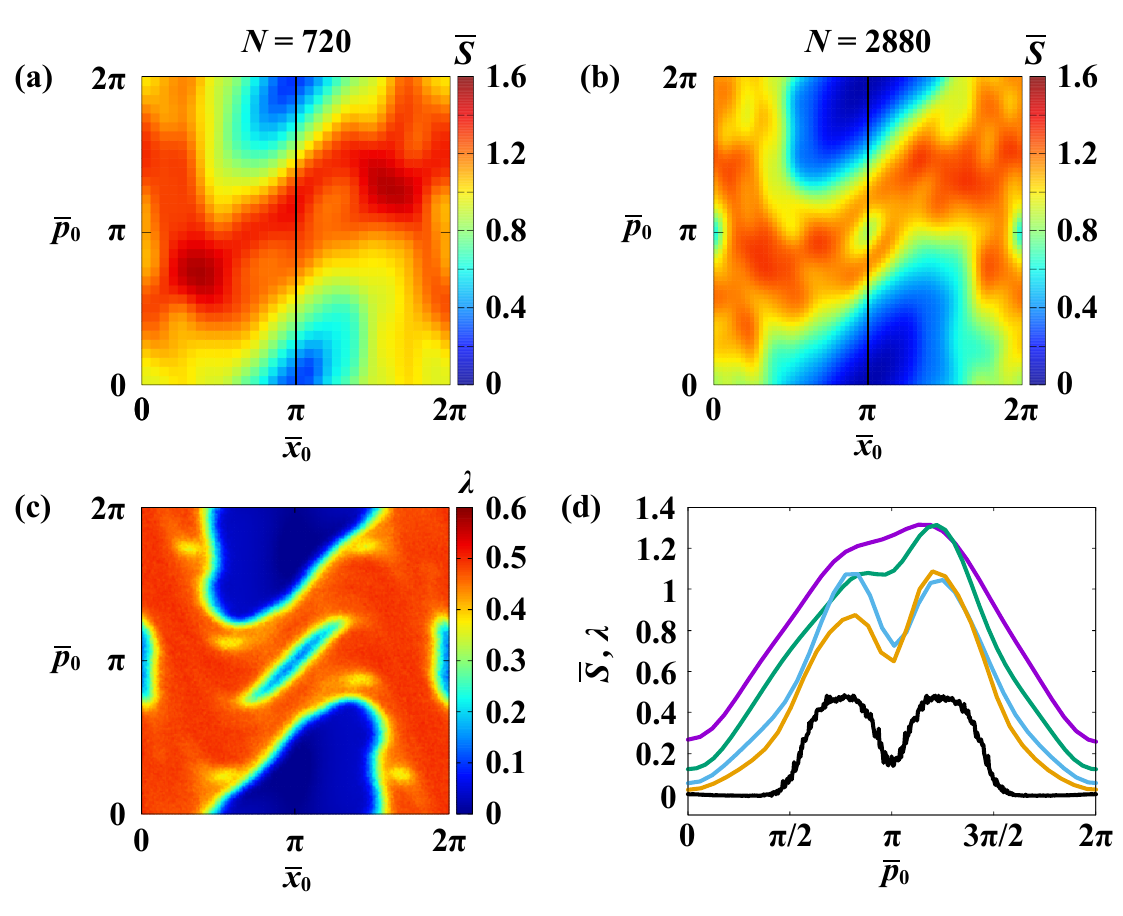}
	\caption{(a), (b) Long-time average $\bar{S}$ of the IEE as a function of the position and momentum of the initial wave packet with (a) $N=720$ and (b) $N=2880$.
		For both cases, the size of the block is $b=40$ and the kick strength is $K=2$.
		(c) Largest Lyapunov exponent $\lambda$ for the classical kicked rotor (calculated over 100 kicks) as a function of the initial points with $K=2$.
		(d) $\bar{S}$ along solid lines shown in (a) and (b) with $N=720$, $1440$, $2880$, $5760$ from top to bottom.
		The lower black line represents $\lambda$.}
	\label{fig-EE-phase-space}
\end{figure}

We calculate the IEE $S(t)$ starting from the initial state given by Eqs.~(\ref{initial_psi})--(\ref{initial_psi_en}).
We first consider the case in which the long-time average is taken before the Hamiltonian equation limit.
In particular, we define the long-time average of the IEE
\begin{equation}
	\bar{S} = \lim_{T \to \infty} \frac{1}{T} \sum_{t=1}^T S(t),
	\label{def_S_bar}
\end{equation}
which is equal to the saturation value of $S(t)$ at long time.
Figures \ref{fig-EE-phase-space} (a) and (b) show $\bar{S}$ as a function of the position $\bar{x}_0$ and momentum $\bar{p}_0$ of the initial wave packet with $K=2$.
The numbers of sites are (a) $N=720$ and (b) $N=2880$, and the block size is $b=40$ for both cases.
Figure \ref{fig-EE-phase-space} (c) shows the largest Lyapunov exponent $\lambda$ for the classical kicked rotor as a function of the initial points.
In order to reduce fluctuations in the data, the largest Lyapunov exponents are averaged over $100$ trajectories with initial points sampled from a Gaussian distribution with mean $(\bar{x}_0, \bar{p}_0)$ and standard deviations $\Delta_x$ and $\Delta_p$.
Here, $\Delta_x$ and $\Delta_p$ are taken to be the same values as the standard deviations $\sigma_x$ and $\sigma_p$ of the wave function (\ref{initial_psi})--(\ref{initial_psi_en}) for $N=2880$.
In Fig.~\ref{fig-EE-phase-space} (d), $\bar{S}$ is plotted as a function of $\bar{p}_0$ with $\bar{x}_0=\pi$ for $N=720$, $1440$, $2880$, and $5760$.
The lower black line represents the largest Lyapunov exponent $\lambda$ as a function of $\bar{p}_0$ with $\bar{x}_0=\pi$.
One can see that $\bar{S}$ closely correlates with $\lambda$ for the classical kicked rotor.
As $N$ increases, the contrast between regular and chaotic regions becomes sharper.

\begin{figure}
	\centering
	\includegraphics[width=0.45\textwidth]{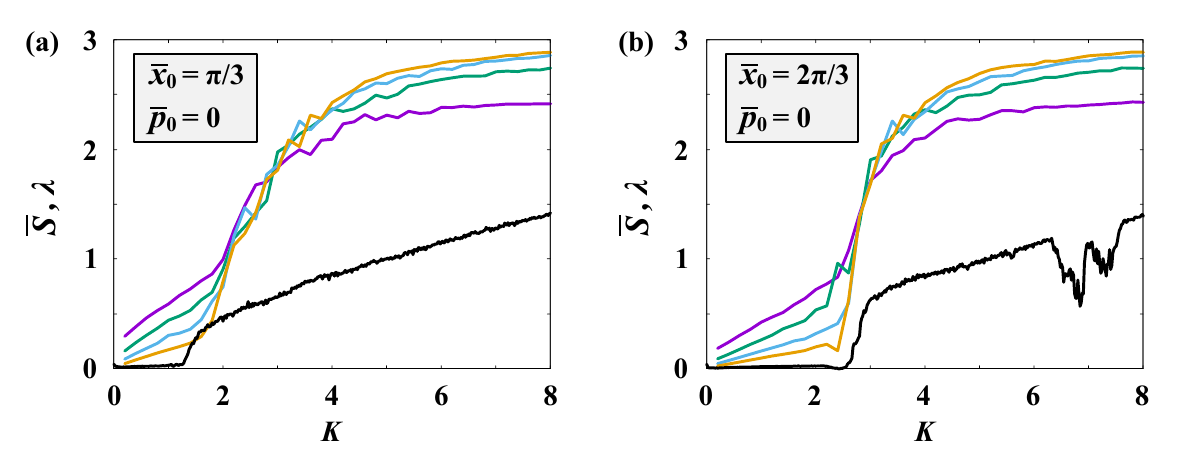}
	\caption{Long-time average $\bar{S}$ of the IEE as a function of the kick strength $K$ with $N=720$, $1440$, $2880$, $5760$ from top to bottom at $K=1$.
		For all cases, the block size is $b=40$.
		The position and momentum of the initial wave packet are (a) $\bar{x}_0=\pi/3$, $\bar{p}_0=0$ and (b) $\bar{x}_0=2\pi/3$, $\bar{p}_0=0$.
		The lower black line represents the largest Lyapunov exponent $\lambda$ (calculated over 100 kicks) as a function of $K$.}
	\label{fig-EE-av}
\end{figure}

Figure \ref{fig-EE-av} shows $\bar{S}$ as a function of $K$ for a fixed initial wave packet, whose position and momentum are (a) $\bar{x}_0=\pi/3$, $\bar{p}_0=0$ and (b) $\bar{x}_0=2\pi/3$, $\bar{p}_0=0$.
The numbers of sites are $N=720$, $1440$, $2880$, and $5760$, and the block size is $b=40$ for all cases.
One can observe a critical kick strength $K_c$ below which $\bar{S}$ vanishes in the limit $N \to \infty$.
For $K>K_c$, $\bar{S}$ converges to a nonzero value.
From Figs.~\ref{fig-EE-av} (a) and (b), $K_c$ can be roughly estimated as $K_c \simeq 1.5$ for $\bar{x}_0=\pi/3$, $\bar{p}_0=0$, and $K_c \simeq 2.5$ for $\bar{x}_0=2\pi/3$, $\bar{p}_0=0$.
In Fig.~\ref{fig-EE-av}, the lower black line represents the largest Lyapunov exponent $\lambda$ for the classical kicked rotor as a function of $K$ for a fixed initial condition: (a) $\bar{x}_0=\pi/3$, $\bar{p}_0=0$ and (b) $\bar{x}_0=2\pi/3$, $\bar{p}_0=0$.
The same averaging procedure as Fig.~\ref{fig-EE-phase-space} (c) is used in the calculation of $\lambda$. 
A dip around $K=6 \sim 7.5$ in Fig.~\ref{fig-EE-av} (b) indicates that the trajectory is trapped in a small regular region.
Let us denote the minimal kick strength for which the trajectory starting from a given initial condition $(\bar{x}_0, \bar{p}_0)$ exhibits chaos as $K_c^{(\mathrm{cl})}(\bar{x}_0, \bar{p}_0)$.
From Fig.~\ref{fig-EE-av}, we have $K_c^{(\mathrm{cl})}(\pi/3,0) \simeq 1.3$ and $K_c^{(\mathrm{cl})}(2\pi/3,0) \simeq 2.8$, which are close to $K_c$ obtained from $\bar{S}$.
These observations imply that the long-time averaged IEE plays a role of an indicator that distinguishes regular and chaotic dynamics.
This result can be understood as follows.
For the regular cases, the microscopic and macroscopic dynamics of the wave packet are decoupled, in other words, its microscopic structure hardly affects the time evolution of the expectation values of the position and momentum.
Thus, the IEE vanishes.
In contrast, for the chaotic cases, since the dynamics at different scales are strongly correlated, the IEE does not vanish.
While $\bar{S}$ can be a qualitative indicator of chaos, the quantitative relation between $\bar{S}$ and $\lambda$ is not clear at present.

We here make some remarks on related previous studies.
The long-time average of entanglement entropy in quantum chaotic systems has long been studied as an indicator of chaos \cite{Furuya-98, Lakshminarayan-01, Bandyopadhyay-04, Ghose-04, Trail-08, Chung-09, Lombardi-11, Casati-12, Matsui-16, Neill-16, Ruebeck-17, Dogra-19}.
In general, for initial wave packets belonging to a chaotic region in classical phase space, the long-time average of entanglement entropy is larger than that of the regular case.
However, note that even for systems with regular behavior, the long-time average of entanglement entropy takes a nonzero value in the classical limit \cite{Casati-12}.
In other words, the standard entanglement entropy between subsystems cannot be a sharp indicator for the onset of chaos.
Interesting exceptional cases are discussed in Refs.~\cite{Wang-04, Ghose-08, Piga-19}, where a single quantum kicked top is considered as a fully-connected multispin system.
In these studies, it is argued that in the limit of an infinite number of spins, the long-time average of the entanglement entropy between one spin and the others is nonzero for chaotic initial states but zero for regular initial states.
This is clearly different from the standard situation of considering entanglement between subsystems with a well-defined classical limit.

\subsection{Entropy production rate}
\label{sec:entropy_production}

\begin{figure}
	\centering
	\includegraphics[width=0.45\textwidth]{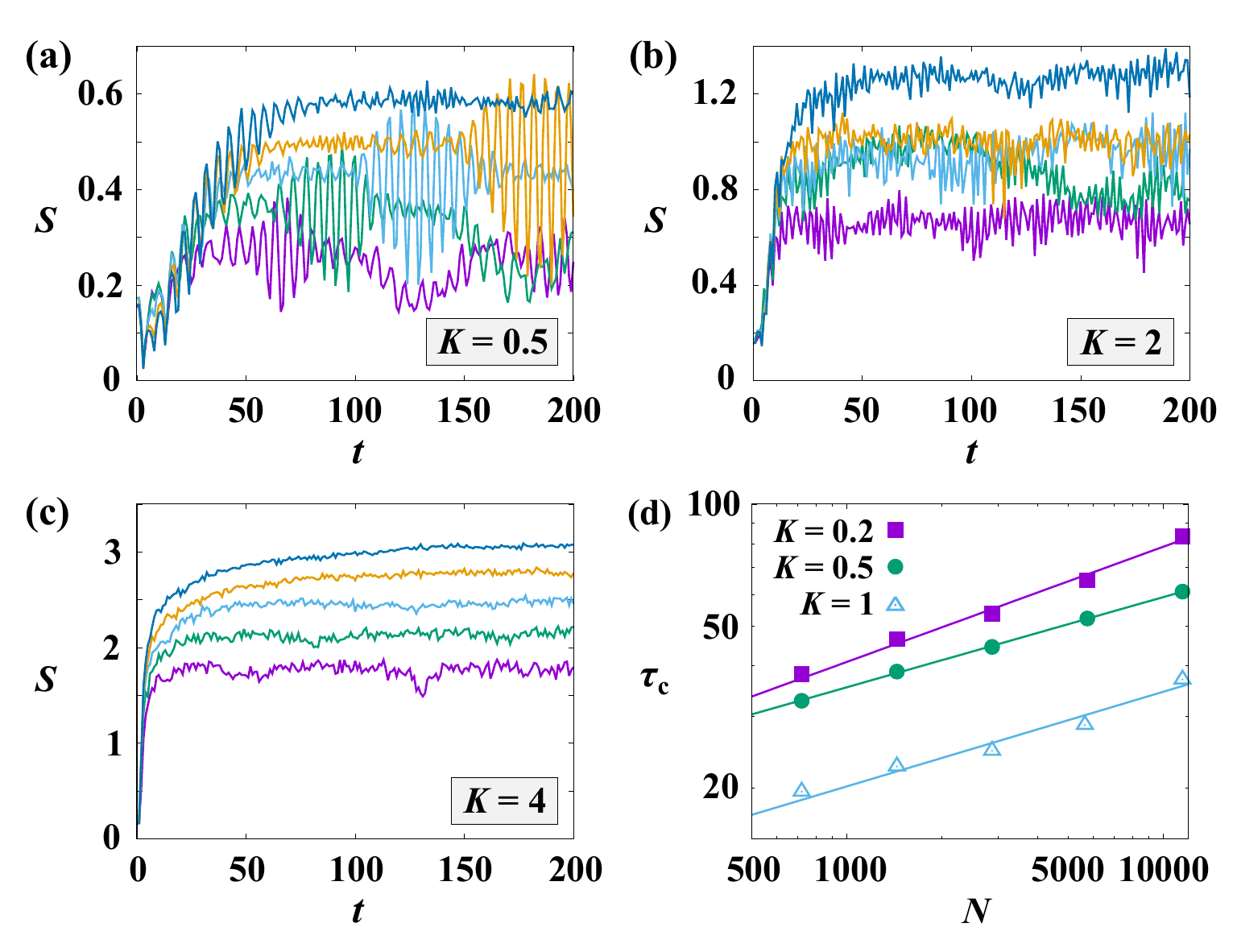}
	\caption{(a)-(c) Time evolution of the IEE $S(t)$ with $(N,b)=$ $(360,20)$, $(720,36)$, $(1440,60)$, $(2880,96)$, $(5760,160)$; from bottom to top.
		The values of the kick strength are (a) $K=0.5$, (b) $K=2$, and (c) $K=4$.
		The initial position and momentum are $\bar{x}_0=\pi/2$ and $\bar{p}_0=0$.
		(d) Crossover timescale $\tau_c$ at which $S(t)$ saturates to an equilibrium value.
		The values of the kick strength are $0.2$, $0.5$, and $1$ from top to bottom.
		The abscissa and ordinate are shown in log scales.
		The straight lines denote the least squares fitting by $\tau_c = A N^{\alpha}$.
		The values of the exponent are estimated as $\alpha=0.28 \ (K=0.2)$, $\alpha=0.22 \ (K=0.5)$, and $\alpha=0.23 \ (K=1)$.}
	\label{fig-EE-dynamics}
\end{figure}

Next, we consider the case in which the Hamiltonian equation limit is take  before the long-time limit.
Figures \ref{fig-EE-dynamics} (a)--(c) show the time evolution of $S(t)$ for different numbers of sites $N$.
The number of the sites $b$ in each block is chosen so that the block size is nearly equal to the width of the initial wave packet: $b/N \simeq \sigma_{x0}/2\pi$.
Since $\sigma_{x0} \propto N^{-1/4}$ from Eq.~(\ref{sigma_x0}), the block size and the number of the blocks increase as $b \propto N^{3/4}$ and $N_b \propto N^{1/4}$, respectively.
As mentioned in Sec.~\ref{sec:interscale_entanglement_entropy}, when the wave function localizes in a single block, the IEE has a small value.
The IEE increases with the expansion of the wave packet, and it finally saturates to some equilibrium value.

For the case $K=0.5$, where the corresponding classical dynamics are regular, $S(t)$ first increases linearly with time and finally saturates at $t=\tau_c$.
Although $S(t)$ shows an oscillating behavior, the mean growth rate in the early stage is almost independent of $N$.
The timescale $\tau_c$ diverges with $N$.
Figure \ref{fig-EE-dynamics} (d) shows $\tau_c$ as a function of $N$ for different values of $K$.
We estimate $\tau_c$ from the cross point between two lines obtained by fitting $S(t)$ in the early and later regimes.
We have $\tau_c \propto N^{\alpha}$ with $\alpha \simeq 1/4$.
Since the width of the initial wave packet scales as $\sigma_{x0} \propto N^{-1/4}$, we expect $\tau_c \propto \sigma_{x0}^{-1}$, which means that the width of the wave packet increases linearly with time.
This result is consistent with the linear dependence of the classical regular trajectory on initial conditions.
For the case $K=4$, where the corresponding classical dynamics are chaotic, $S(t)$ shows a rapid saturation within a short timescale and does not show any oscillating behavior except for irregular fluctuations.
The saturation value of $S(t)$ is much larger than that for the regular case.
Although $\tau_c$ is too short to estimate its $N$-dependence, it is assumed to be given by Eq.~(\ref{tau_c}).
If the Lyapunov exponent is $O(1)$, $\tau_c$ becomes only several cycles of the kick even for $N=5760$.
For the case $K=2$, where the classical phase space is filled with both regular and chaotic trajectories almost equally (see Fig.~\ref{fig-EE-phase-space} (c)), one can observe an intermediate behavior between the two cases mentioned above.

Let us consider the production rate $h_{\mathrm{Q}}$ of the IEE.
For the regular case ($K<2$), $h_{\mathrm{Q}}$ can be evaluated from the mean slope of $S(t)$ in the early stage.
For the chaotic case ($K>4$), the crossover timescale $\tau_c$ is too small to evaluate $h_{\mathrm{Q}}$ by the linear fitting.
Thus, it is convenient to define $h_{\mathrm{Q}}$ by
\begin{equation}
	h_{\mathrm{Q}} = \frac{S(t=3)-S(t=0)}{3},
\end{equation}
which measures a rapid increment of $S(t)$ in the first three kicks (see Fig.~\ref{fig-EE-dynamics} (c)).
The question is how the Hamiltonian equation limit of $h_{\mathrm{Q}}$
\begin{equation}
	h_{\mathrm{Q}}^{\infty} = \lim_{N \to \infty} h_{\mathrm{Q}}
\end{equation}
is related to a quantity of the classical dynamics.

\begin{figure}
	\centering
	\includegraphics[width=0.45\textwidth]{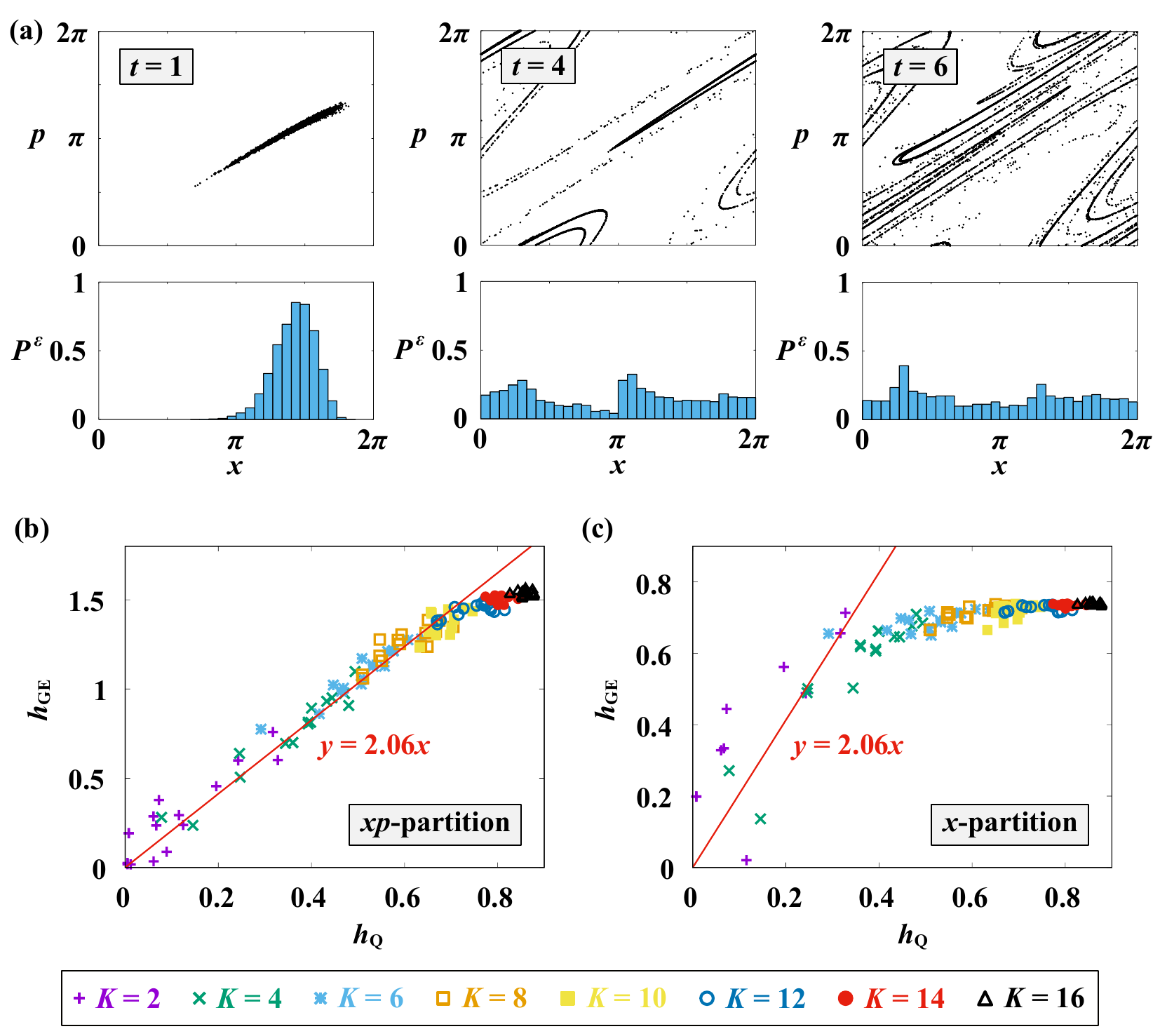}
	\caption{(a) Time evolution of the phase space distribution for the classical kicked rotor with $K=4$.
		The initial distribution is a Gaussian distribution with mean $\bar{x}_0=\pi/3$, $\bar{p}_0=0$ and standard deviations $\Delta_x=0.154$, $\Delta_p=0.092$, which are equal to the standard deviations of the wave function (\ref{initial_psi})--(\ref{initial_psi_en}) for $N=2880$.
		The lower panels show the coarse-grained distribution $P_{\alpha}^{\epsilon}$ for the $x$-partition with $\epsilon_x=2\pi/30$.
		(b), (c) $h_{\mathrm{Q}}$ versus $h_{\mathrm{GE}}$ for $K=2$, $4$, $6$, $8$, $10$, $12$, $14$, $16$, and different initial conditions.
		For each value of $K$, $16$ initial conditions $(x_0, p_0)$ are uniformly sampled from $[0,2\pi] \times [0,\pi]$.
		The IEE is calculated for $N=2880$ and $b=96$.
		The coarse-grained Gibbs entropy is calculated for (b) the $xp$-partition and (c) the $x$-partition of the phase space.
		The solid lines show a linear fitting of data for the $xp$-partition up to $K=10$ by $h_{\mathrm{GE}}=\kappa h_{\mathrm{Q}}$ with $\kappa=2.06$ , and the relative standard error of $\kappa$ is $0.8\%$.}
	\label{fig-h_Q-h_GE}
\end{figure}

We consider the relationship between $h_{\mathrm{Q}}^{\infty}$ and the production rate of the coarse-grained Gibbs entropy $S_{\mathrm{cl}}^{\epsilon}(t)$ for the classical dynamics, which is defined in Appendix~\ref{appendix:Gibbs_entropy}.
Here, the entropy $S_{\mathrm{cl}}^{\epsilon}(t)$ depends on the choice of the partition $\{ \Lambda_{\alpha}^{\epsilon} \}_{\alpha=1,2,...}$ of the phase space.
We then consider two types of partition, the ``$x$-partition'' and the ``$xp$-partition''.
In the former case, the phase space is divided into stripe-shaped cells parallel to the $p$-direction.
The width $\epsilon_x$ of the cells along the $x$-direction is set to the same value as the block size in the calculation of the IEE, namely $\epsilon_x = 2\pi b/N$.
On the other hand, in the $xp$-partition, the phase space is divided into rectangular-shaped cells.
The width $\epsilon_x$ of the cells along the $x$-direction is given by $\epsilon_x = 2\pi b/N$ and the width $\epsilon_p$ along the $p$-direction is determined by the uncertainty relation \eqref{appendix:uncertainty}, namely $\epsilon_p = Ja^2/\epsilon_x$.
The phase space distribution of initial conditions is chosen as a Gaussian distribution that has the same mean values $\bar{x}_0$, $\bar{p}_0$ and standard deviations $\sigma_x$, $\sigma_p$ as the wave function given by Eqs.~(\ref{initial_psi})--(\ref{initial_psi_en}).
The numerical method for the calculation of the coarse-grained Gibbs entropy is explained in Appendix~\ref{appendix:Gibbs_entropy}.
To restrict the phase space distribution in $[0,2\pi) \times [0,2\pi)$, we impose a periodic boundary condition with respect to the momentum: $p+2n\pi \to p$.
We consider the growth rate of $S_{\mathrm{cl}}^{\epsilon}(t)$ in the early stage,
\begin{equation}
	h_{\mathrm{GE}} = \frac{S_{\mathrm{cl}}^{\epsilon}(t=3)-S_{\mathrm{cl}}^{\epsilon}(t=0)}{3}.
	\label{h_GE_def}
\end{equation}

Figure \ref{fig-h_Q-h_GE} (a) shows the time evolution of the phase space distribution for the classical kicked rotor.
The kick strength is set to $K=4$, in which case most of the phase space is filled with chaotic trajectories.
We can see ``mixing" of the phase space distribution by repeated stretching and folding.
The lower panels of Fig.~\ref{fig-h_Q-h_GE} (a) show the coarse-grained distribution $P_{\alpha}^{\epsilon}$ for the $x$-partition, which is defined by Eq.~\eqref{P_epsilon}.
The delocalization of the coarse-grained distribution results in the increase of the Gibbs entropy $S_{\mathrm{cl}}^{\epsilon}$.

Figures \ref{fig-h_Q-h_GE} (b) and (c) show $h_{\mathrm{Q}}$ versus $h_{\mathrm{GE}}$ for different values of $K$ and several initial conditions.
For the case of the $xp$-partition (b), while the values of $h_{\mathrm{Q}}$ and $h_{\mathrm{GE}}$ fluctuate depending on the initial conditions, one can see a linear relationship
\begin{equation}
	h_{\mathrm{GE}} \simeq \kappa h_{\mathrm{Q}}
	\label{h_Q_h_E}
\end{equation}
with $\kappa \simeq 2.06$.
The value of $\kappa$ is found to be almost independent of $N$ as long as $N$ is sufficiently large.
For large $K$, $h_{\mathrm{GE}}$ saturates and deviates from the linear relationship \eqref{h_Q_h_E}.
This is because the linear growth regime of $S_{\mathrm{cl}}^{\epsilon}(t)$ in early time becomes narrower as $K$ increases.
When the interval of the linear growth regime is shorter than three kicks, $h_{\mathrm{GE}}$ given by Eq.~\eqref{h_GE_def} starts to saturate.
Note that the interval of the linear growth regime increases as the width of the partition $\epsilon_{x,p}$ and that of the initial distribution $\sigma_{x,p}$ go to zero.
Thus, we expect that the linear behavior observed in Fig.~\ref{fig-h_Q-h_GE} (b) persists for any large $K$ in the limit $N \to \infty$.
For the case of the $x$-partition (c), $h_{\mathrm{GE}}$ quickly saturates to a constant value, and the linear relationship \eqref{h_Q_h_E} cannot be observed clearly.
This is because when $K$ is large, the coarse-grained distribution for the $x$-partition spreads to the whole position space very early.
In fact, Fig.~\ref{fig-h_Q-h_GE} (a) shows that the coarse-grained distributions at $t=4$ and $t=6$ have already spread over the whole space and the Gibbs entropy has reached its maximum, but the full distribution has not yet covered the whole phase space.

There have been many studies on the dynamical generation of entanglement in quantum systems that exhibit chaos in the classical limit \cite{Zurek-94, Zarum-98, Miller-99-1, Miller-99-2, Pattanayak-99, Monteoliva-00, Monteoliva-01, Demkowicz-04, Asplund-16, Bianchi-18, Tanaka-02, Fujisaki-03, Jacquod-04, Jacquod-09, Furuya-98, Lakshminarayan-01, Bandyopadhyay-04, Ghose-04, Trail-08, Chung-09, Lombardi-11, Casati-12, Matsui-16, Neill-16, Ruebeck-17, Dogra-19}.
For example, in Ref.~\cite{Casati-12}, the generation of entanglement between two interacting particles is investigated.
The entanglement entropy increases linearly with time and then saturates to a certain value.
The timescale of the linear growth regime is given by $\sqrt{A/\hbar}$ when the dynamics of the corresponding classical system is regular, and by $\log(A/\hbar)$ when it is chaotic, where $A$ is a typical classical action.
Moreover, the time evolution of the entanglement entropy is found to be well approximated by the classical coarse-grained entropy in the semiclassical regime $A/\hbar \gg 1$.
This behavior of the entanglement between the interacting particles is similar to the behavior of the IEE described above.
Note that in our case, $a^{1/2} \ (\propto N^{-1/2})$ plays the role of the effective Planck's constant (see Eq.~\eqref{h_eff}).

\section{Conclusions}
\label{sec:conclusions}

In this study, we investigated standard classical chaos in terms of entanglement between degrees of freedom at different scales, the interscale entanglement entropy (IEE).
Initially, we designed a quantum lattice system that simulates classical chaos after an appropriate continuum limit, the Hamiltonian equation limit.
For such a lattice system, the IEE is defined by the block-spin coarse graining procedure.
We numerically calculated the IEE of the kicked rotor model and found that the long-time average of the IEE takes a finite value when chaos occurs, but the IEE converges to zero when the dynamics is regular.
Furthermore, we show that the initial growth rate of the IEE is strongly correlated with the growth rate of the coarse-grained Gibbs entropy for the corresponding classical dynamics.

The IEE is thought to characterize the spread of the wave packet in phase space rather than in real space.
Even though the wave function is spread over the whole system in both the regular and chaotic cases, the long-time average of the IEE is found to be larger in the latter case.
This implies that the IEE is not only determined by the spread of the wave packet in real space.
The large value of the IEE reflects the fact that in chaotic dynamics, the wave packet has a large spread in the momentum direction.
Moreover, the growth rate of the IEE is more strongly correlated with the Gibbs entropy coarse-grained in phase space than with the Gibbs entropy coarse-grained in real space.
In other words, while the IEE is defined by coarse-graining in real space, it also contains information about the spread of the wave packet in momentum space.

The growth rate of the classical Gibbs entropy is expected to be identical to the KS entropy \cite{Latora-99, Vulpiani-05}.
Our results suggest that the IEE provides a microscopic representation of the KS entropy in terms of quantum entanglement.
In classical chaotic systems, the exponential sensitivities of the trajectories to the initial conditions cause microscopic details in phase space to expand into macroscopic structures.
In other words, the KS entropy can be interpreted as a measure of the information flow from the microscopic scale to the macroscopic scale.
Therefore, it is reasonable to conjecture that the KS entropy is closely related to the production rate of the IEE, which measures the amount of mutual information between microscopic and macroscopic degrees of freedom.
It is a matter of future work to scrutinize this conjecture quantitatively and to provide a theoretical basis for it.

We mention the relationship between the IEE and out-of-time-ordered correlators (OTOC) \cite{Larkin-69, Maldacena-16, Hashimoto-17, Swingle-18}.
In our setup, the OTOC is defined by $C(t) = - \langle [\hat{x}(t), \hat{p}(0)]^2 \rangle$ in terms of $\hat{x}$ and $\hat{p}$ given by Eqs.~\eqref{def_x_hat} and \eqref{def_p_hat}.
The OTOC is a direct measure of the sensitivities of semiclassical trajectories to initial conditions and is expected to grow exponentially with the classical Lyapunov exponent.
In classical chaotic systems, there is a close relationship between the exponential separation of close trajectories and the production of the entropy (see Eq.~\eqref{appendix:Pesin} in Appendix \ref{appendix:KS_entropy}).
Thus, the initial time behavior of the IEE $S(t)$ and $\log C(t)$ is expected to be qualitatively similar.
However, the long-time averages of these quantities can exhibit quite different behavior.
In particular, even when the dynamics is regular, the long-time average of the OTOC would not vanish in the classical limit.

Finally, we discuss possible extensions of this study.
It would be interesting to investigate whether the results presented in this work hold true for dissipative systems coupled to an environment.
The definition of the IEE can be extended directly to mixed states described by a density matrix.
Thus, it is possible to compare the IEE with the classical Gibbs entropy for a dissipative chaotic system, e.g., a kicked rotor subjected to a dissipative friction force discussed in Ref.~\cite{Carlo-05}.
We expect that even in the presence of dissipation, there is still a strong correlation between the production rate of the IEE and that of the classical Gibbs entropy.
In another direction, extending the concept of the IEE to many-body systems will be an interesting challenge in the future.
For example, the dynamics of the Bose-Hubbard model on a lattice is described by the path integral formalism in terms of the boson coherent states.
By discretizing the phase space of the boson coherent states, it may be possible to construct a quantum model (``unified simulator") with two continuum limits that reproduce the original quantum model and the classical model described by the Gross-Pitaevskii equation.
Since numerical simulation of such a model is a formidable task, it would be reasonable to start with a model with fewer degrees of freedom, e.g., two-site Bose-Hubbard model or the Dicke model.

\begin{acknowledgments}
This study was supported by JSPS KAKENHI Grant Numbers JP17H01148, JP19J00525, JP19H05795, and JP20K20425.
\end{acknowledgments}

\appendix
\section{Kolmogorov--Sinai entropy}
\label{appendix:KS_entropy}

\begin{figure}
	\centering
	\includegraphics[width=0.3\textwidth]{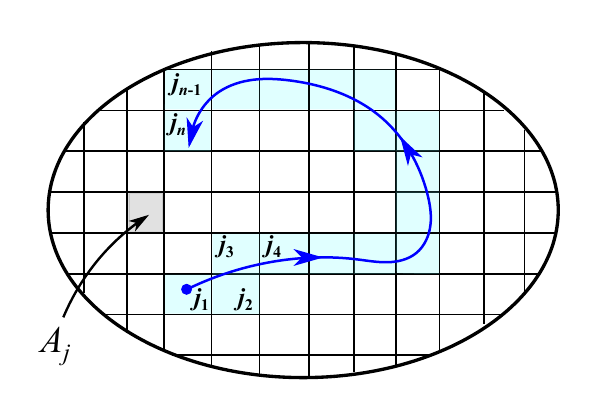}
	\caption{Schematic illustration of the trajectory in the partitioned phase space.
		$\mathcal{A}=\{A_1,...,A_r \}$ is a finite partition of the phase space.
		A continuous trajectory $f_t(\Gamma)$ generates a series of words $\{ j_1,...,j_n \}$.}
	\label{fig-discrete-trajectory}
\end{figure}

In this section, we provide a brief overview of the basic concepts \cite{Eckmann-85, Boffetta-02}.
We denote a state in the phase space $\mathcal{X}$ as $\Gamma$, and its time evolution during time $t$ as $\Gamma \to f_t(\Gamma)$.
The invariant measure $\mu$ on the phase space satisfies $\mu(f_t(A))=\mu(A)$ for all subsets $A \subset \mathcal{X}$ and any $t$.
Let $\mathcal{A}=\{A_1,...,A_r \}$ be a finite partition of the phase space, namely, $A_i \cap A_j = \emptyset$ for all $i \neq j$ and $\mu(\mathcal{X}-\cup_{i=1}^r A_r)=0$.
For a given $\Gamma$, a discrete trajectory $\{ j_1,...,j_n \}$ is generated from the condition $f_{t=kT}(\Gamma) \in A_{j_k}$ for some time interval $T$ (see Fig.~\ref{fig-discrete-trajectory}).
The probability distribution associated with the trajectory $\{ j_1,...,j_n \}$ is defined by
\begin{equation}
	p(j_1,...,j_n) = \mu(f_{t=T}^{-1}(A_{j_1}) \cap ... \cap f_{t=nT}^{-1}(A_{j_n})),
\end{equation}
where $f_t^{-1}(A)$ is the set obtained by evolving each point in $A$ backwards during $t$.
The corresponding Shannon entropy reads
\begin{equation}
	H_n(\mathcal{A}) = - \sum_{j_1,...,j_n} p(j_1,...,j_n) \ln p(j_1,...,j_n).
\end{equation}
The KS entropy $h_{\mathrm{KS}}$ is defined by
\begin{equation}
	h_{\mathrm{KS}} = \sup_{\mathcal{A}} \lim_{n \to \infty} \frac{H_n(\mathcal{A})}{n T} ,
	\label{appendix:def_h_KS}
\end{equation}
where $\sup$ is taken over all finite partitions $\mathcal{A}$.
Note that $h_{\mathrm{KS}}$ is independent of the time interval $T$.
The quantity $h_{\mathrm{KS}}$ can be interpreted as the exponentially growing rate of the number of the typical trajectories.

It is almost impossible to numerically estimate $h_{\mathrm{KS}}$ according to its original definition (\ref{appendix:def_h_KS}).
However, the following formula provides us a useful alternative way to calculate it in terms of the Lyapunov exponents $\lambda_i$, which can be easily calculated numerically,
\begin{equation}
	h_{\mathrm{KS}} = \sum_{\lambda_i>0} \lambda_i,
	\label{appendix:Pesin}
\end{equation}
where the summation is taken over all positive Lyapunov exponents.
This formula, which is called the Pesin theorem, implies that $h_{\mathrm{KS}}$ is closely related to the exponentially growing rate of the separation between nearby trajectories.

\section{Coarse-grained Gibbs entropy}
\label{appendix:Gibbs_entropy}

We discuss the connection between $h_{\mathrm{KS}}$ and the growth rate of the coarse-grained Gibbs entropy.
Let $\rho(\Gamma;t)$ be the probability density function on the phase space at time $t$.
The Gibbs entropy is defined as
\begin{equation}
	S_{\mathrm{cl}}(t) = - \int d \Gamma \rho(\Gamma;t) \ln \rho(\Gamma;t).
	\label{Gibbs_entropy}
\end{equation}
In the case of a volume-conserving evolution, the Gibbs entropy does not change, $S_{\mathrm{cl}}(t)=S_{\mathrm{cl}}(0)$.

To define a coarse-grained Gibbs entropy, we partition the phase space into small cells $\{ \Lambda_{\alpha}^{\epsilon} \}_{\alpha=1,2,...}$ with length $\epsilon$.
The coarse-grained probability density is defined as
\begin{equation}
	P_{\alpha}^{\epsilon}(t) = \int_{\Lambda_{\alpha}^{\epsilon}} d \Gamma \rho(\Gamma;t),
	\label{P_epsilon}
\end{equation}
and the coarse-grained Gibbs entropy reads
\begin{equation}
	S_{\mathrm{cl}}^{\epsilon}(t) = - \sum_{\alpha} P_{\alpha}^{\epsilon}(t) \ln P_{\alpha}^{\epsilon}(t).
\end{equation}

Starting from a localized initial distribution $\rho(\Gamma;0)$, whose width is comparable to $\epsilon$, $S_{\mathrm{cl}}^{\epsilon}(t)$ grows linearly in time until it saturates to an equilibrium value.
It is conjectured that the initial growth rate of $S_{\mathrm{cl}}^{\epsilon}(t)$ is equal to the KS entropy \cite{Latora-99,Vulpiani-05}:
\begin{equation}
	S_{\mathrm{cl}}^{\epsilon}(t) - S_{\mathrm{cl}}^{\epsilon}(0) \simeq h_{\mathrm{KS}} t.
\end{equation}

We briefly mention the numerical method to calculate the coarse-grained Gibbs entropy.
First, an ensemble of initial conditions $\{ \Gamma_i(0) \}_{i=1,...,M}$ is randomly sampled from a given initial distribution $\rho(\Gamma;0)$.
Then, the time evolution of $\{ \Gamma_i(t) \}_{i=1,...,M}$ is obtained by numerically integrating the equation of motion.
Let $N_{\alpha}(t)$ be the number of $\{ \Gamma_i(t) \}_{i=1,...,M}$ that belongs to the $\alpha$ th cell $\Lambda_{\alpha}^{\epsilon}$, which satisfies $\sum_{\alpha} N_{\alpha}(t) = M$.
The coarse-grained Gibbs entropy is then given by
\begin{equation}
	S_{\mathrm{cl}}^{\epsilon}(t) = - \sum_{\alpha} \frac{N_{\alpha}(t)}{M} \ln \frac{N_{\alpha}(t)}{M}.
\end{equation}
In numerical calculation in Sec.~\ref{sec:entropy_production}, the size of the ensemble $M$ is set to $10000$.

\section{Hamiltonian equation limit}
\label{appendix:Hamiltonian_equation_limit}

In this Appendix, we discuss the derivation of the Hamiltonian equation limit in the tight-binding model.
In the Heisenberg representation, the time evolution of the position operator \eqref{def_x_hat} is given by
\begin{equation}
	\frac{d \hat{x}}{dt} = \frac{i}{\hbar}[\hat{H},\hat{x}] = i\frac{Ja}{\hbar} \sum_n (\hat{a}_{n+1}^{\dag} \hat{a}_n - \hat{a}_{n}^{\dag} \hat{a}_{n+1}).
\end{equation}
Thus, Eq.~\eqref{x_EOM} follows from the definition of the momentum operator \eqref{def_p_hat}.
The time evolution of $\hat{p}$ is given by
\begin{align}
	\frac{d \hat{p}}{dt} &= \frac{mJga}{\hbar^2} \sum_n [U(x_n)-U(x_{n+1})] (\hat{a}_{n+1}^{\dag} \hat{a}_n + \hat{a}_{n}^{\dag} \hat{a}_{n+1}) \notag \\
	&\simeq - \frac{mJga^2}{\hbar^2} \sum_n U'(x_n) (\hat{a}_{n+1}^{\dag} \hat{a}_n + \hat{a}_{n}^{\dag} \hat{a}_{n+1}),
\end{align}
where we have assumed the condition $a \ll l_U$ from Eq.~\eqref{classical_condition_1}.
Furthermore, since the width of the wave packet $\sigma_x$ is much smaller than $l_U$, we have
\begin{eqnarray}
	\frac{d \langle \hat{p} \rangle}{dt}  \simeq -\frac{mJga^2}{\hbar^2} U'(\langle \hat{x} \rangle) \sum_n \langle \hat{a}_{n+1}^{\dag} \hat{a}_n + \hat{a}_{n}^{\dag} \hat{a}_{n+1}\rangle.
	\label{appendix:p_EOM_1}
\end{eqnarray}
The expectation value in the right-hand side of Eq.~(\ref{appendix:p_EOM_1}) can be rewritten as
\begin{align}
	\sum_n \langle \hat{a}_{n+1}^{\dag} \hat{a}_n + \hat{a}_{n}^{\dag} \hat{a}_{n+1} \rangle &= \sum_n  2 |\psi_{n+1}| |\psi_n| \cos(\theta_{n+1}-\theta_n) \notag \\
	&\simeq \sum_n 2 |\psi_n|^2 = 2.
	\label{appendix:appro_kinetic}
\end{align}
where $\psi_n=|\psi_n|e^{i\theta_n}$ and we have assumed the condition \eqref{classical_condition_2}.
Thus, we have the equation of motion \eqref{p_EOM}.

We next determine the condition of the scaling exponent $\beta$ in Eq.~\eqref{J_g_scaling}.
First, we consider the condition (\ref{classical_condition_2}).
The expectation value of the momentum is written as
\begin{equation}
	\langle \hat{p} \rangle = \frac{2mJa}{\hbar} \sum_n |\psi_{n+1}| |\psi_n| \sin( \theta_{n+1} - \theta_{n})
	\simeq \frac{2mJa}{\hbar} \Delta \theta,
\end{equation}
where $\Delta \theta$ represents the averaged phase difference between adjacent sites.
Since $\langle \hat{p} \rangle$ should be independent of $a$, in order for $\Delta \theta$ to vanish in the limit $a \to 0$, $\beta$ must be positive.
Next, we determine the range of $\beta$ which ensures Eq.~(\ref{classical_condition_1}).
Note that, as the standard deviation of the momentum $\sigma_p := (\langle \hat{p}^2 \rangle - \langle \hat{p} \rangle^2)^{1/2}$ for the initial state becomes larger, the timescale $\tau_c$ in which $\sigma_x$ becomes comparable to $l_U$ decreases.
Thus, in order for $\tau_c$ to diverge to infinity in the limit $a \to 0$, $\sigma_x$ and $\sigma_p$ must simultaneously vanish.
The uncertainty relation between the position and momentum is given by
\begin{equation}
	\sigma_x \sigma_p \geq \frac{1}{2} |\langle [\hat{x},\hat{p}] \rangle|.
\end{equation}
The commutator between the position and momentum operators can be calculated as
\begin{equation}
	[\hat{x},\hat{p}] = i \frac{mJa^2}{\hbar} \sum_n (\hat{a}_{n+1}^{\dag} \hat{a}_n + \hat{a}_{n}^{\dag} \hat{a}_{n+1}).
\end{equation}
By using Eq.~(\ref{appendix:appro_kinetic}), we have
\begin{equation}
	\sigma_x \sigma_p \geq \frac{mJa^2}{\hbar}.
	\label{appendix:uncertainty}
\end{equation}
In the case of the Schr\"odinger equation limit, from $Ja^2=\hbar^2/2m$ one can confirm that Eq.~(\ref{appendix:uncertainty}) reduces to the conventional uncertainty relation $\sigma_x \sigma_p \geq \hbar/2$.
If $\beta<1$, the right-hand side of Eq.~(\ref{appendix:uncertainty}) vanishes in the limit $a \to 0$.
By combining two conditions, we obtain Eq.~\eqref{beta_CCL}.

Let us estimate $\tau_c$ as a function of $a$.
If we choose an initial wave packet with the minimal uncertainty, both $\sigma_x$ and $\sigma_p$ are proportional to $a^{(1-\beta)/2}$.
Since the dynamics of the Wigner function associated with the wave function $\psi_n$ is expected to be described by the classical Liouville equation for $t \ll \tau_c$, for chaotic dynamics the time evolution of $\sigma_x$ is given by $\sigma_x(t) \sim \sigma_x(0) e^{\lambda t}$, where $\lambda$ is the largest Lyapunov exponent.
By noting that $\tau_c$ is given by the condition $\sigma_x(\tau_c) \sim l_U$, we have Eq.~\eqref{tau_c}.
In the conventional semiclassical limit of quantum systems, the timescale on which the quantum evolution of a localized wave packet closely follows the corresponding classical trajectory is known as the Ehrenfest time \cite{Berman-78, Berry-79, Silvestrov-02, Schubert-12}, 
\begin{equation}
	\tau_E \sim \frac{1}{\lambda} \ln \frac{A}{\hbar},
\end{equation}
where $A$ is a typical classical action.
As can be seen from the uncertainty relation \eqref{appendix:uncertainty}, 
\begin{equation}
	\hbar_{\mathrm{eff}} := \frac{2mJa^2}{\hbar} \propto a^{1-\beta}
	\label{h_eff}
\end{equation}
plays the role of an effective Planck's constant from the viewpoint of the wave packet dynamics.
Thus, $\tau_c$ given by Eq.~\eqref{tau_c} can be interpreted as the Ehrenfest time associated with $\hbar_{\mathrm{eff}}$.

\section{Cases with different values of $\beta$}
\label{appendix:different_beta}

\begin{figure}
	\centering
	\includegraphics[width=0.45\textwidth]{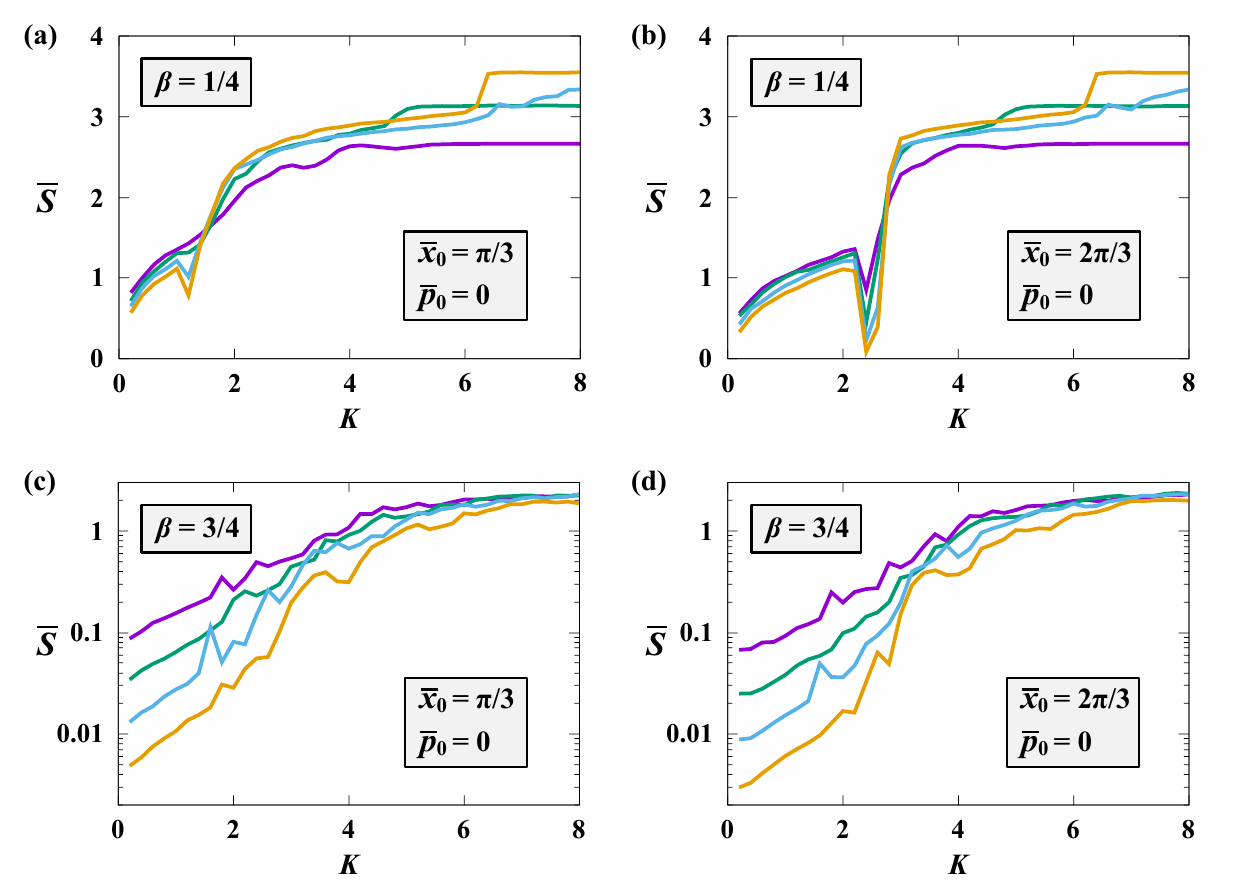}
	\caption{Long-time average $\bar{S}$ of the IEE as a function of the kick strength $K$ with $N=720$, $1440$, $2880$, $5760$ from top to bottom at $K=1$.
		For all cases, the block size is $b=40$.
		The position and momentum of the initial wave packet are $\bar{x}_0=\pi/3$, $\bar{p}_0=0$ for (a) and (c), and $\bar{x}_0=2\pi/3$, $\bar{p}_0=0$ for (b) and (d).
		The values of $\beta$ are $1/4$ (top) and $3/4$ (bottom).
		The vertical axis of (c) and (d) is plotted on a logarithmic scale.}
	\label{fig:EEav_different_beta}
\end{figure}

As discussed in Sec.~\ref{sec:unified_simulator}, to reproduce classical dynamics in the continuum limit $a\to0$, the exponent $\beta$, which controls the scaling of $J$ and $g$ by Eq.~(\ref{J_g_scaling}), must satisfy $0<\beta<1$.
In the numerical calculations presented in Sec.~\ref{sec:kicked_rotor}, $\beta$ is set to $1/2$.
We believe that the qualitative results of this work are independent of the choice of $\beta$, as long as the condition $0<\beta<1$ is satisfied.
For example, in Fig.~\ref{fig:EEav_different_beta}, we show the long-time average $\bar{S}$ of the IEE for $\beta=1/4$ and $3/4$.
The scaling of $J$ and $g$ for general $\beta$ reads
\begin{equation}
	J = a^{-1-\beta}, \:\:\:\: g= \frac{1}{2} a^{-1+\beta}.
	\label{J_g_KR_general_beta}
\end{equation}
The initial state is given by Eq.~(\ref{initial_psi}), where the initial width $\sigma_{x0}$ is taken as
\begin{equation}
	\sigma_{x0} = a^{(1-\beta)/2},
	\label{sigma_x0_general_beta}
\end{equation}
which follows from the uncertainty relation $\sigma_x \sigma_p \sim a^{(1-\beta)}$ (see Eq.~(\ref{appendix:uncertainty}) in Appendix \ref{appendix:Hamiltonian_equation_limit}).
The qualitative behavior of $\bar{S}$ in Fig.~\ref{fig:EEav_different_beta} is the same as in Fig.~\ref{fig-EE-av}.
In fact, $\bar{S}$ is almost independent of $N$ in the chaotic regime, whereas $\bar{S}$ is decreasing with increasing $N$ in the regular regime.
For $\beta=1/4$ (see Figs.~\ref{fig:EEav_different_beta} (a) and (b)), the decrease of $\bar{S}$ with respect to $N$ in the regular regime is more gradual than in the case $\beta=1/2$.
Conversely, the case $\beta=3/4$ shows a faster decrease in $\bar{S}$ in the regular regime than the case $\beta=1/2$ (see Figs.~\ref{fig:EEav_different_beta} (c) and (d)).
Assuming $\bar{S} \sim N^{-\eta}$ in the regular regime, the exponent $\eta$ is estimated to be $\eta=0.11 \pm 0.02$ for $\beta=1/4$, $\eta=0.695 \pm 0.004$ for $\beta=1/2$, and $\eta=1.29 \pm 0.01$ for $\beta=3/4$ by least squares fitting of $\bar{S}$ for $\bar{x}_0=2\pi/3$, $\bar{p}_0=0$ at $K=1$.

\end{document}